\documentclass{ifacconf}
\usepackage{graphicx}
\usepackage{natbib}

\usepackage{amsmath}
\usepackage{amssymb}
\usepackage{mathrsfs}
\usepackage{amsfonts}
\usepackage{color}
\usepackage{textcomp}
\usepackage{autobreak}

\newtheorem{definition}{Definition}

\newtheorem{assumption}{Assumption}

\newtheorem{lemma}{Lemma}
\newtheorem{remark}{Remark}

\usepackage{algpseudocode}

\usepackage{array}

\usepackage{cases}

\usepackage{stfloats}

\usepackage{pbox,pifont}
\usepackage[caption=false,font=normalsize,labelfont=sf,textfont=sf]{subfig}

\usepackage{url}

\begin{document}
\begin{frontmatter}

\title{Consensus Tracking of Perturbed Open Multi-Agent Systems with Repelling Antagonistic Interactions\thanksref{footnoteinfo}}
\thanks[footnoteinfo]{This work was supported by the National Natural Science Foundation of China under Grant 62473296, the Fundamental Research Funds for the Central Universities, and the Shanghai Pilot Program for Basic Research.}

\author[Paestum]{Mengqi Xue}
\author[Paestum]{Yuchao Xiong}
\author[Paestum]{Yue Song}

\address[Paestum]{College of Electronic and
Information Engineering, State Key Laboratory of Autonomous
Intelligent Unmanned Systems, Tongji University, Shanghai 200092, P.R. China (e-mails: x\_starter@hotmail.com; 2431989@tongji.edu.cn; ysong@tongji.edu.cn).}

\begin{abstract}
An open multi-agent system (OMAS) features migrating agents which produce a flexible network that is naturally switching and size-varying. Meanwhile, agent migrations also make an OMAS prone to environmental adversities. In this work, we investigate the consensus tracking problem of OMASs suffering migration-induced adversities, including non-vanishing agent dynamics/state perturbations and repelling antagonistic interactions among agents, over an intermittently disconnected signed digraph. The OMAS is interpreted into a perturbed multi-mode multi-dimensional ($M^3D$) system in which unstable subsystems are created when repelling interactions dominate the cooperative ones in the network regardless of its connectivity. To handle the destabilizing effect brought by repelling interactions and non-vanishing perturbations, we extend the stability theory for $M^3D$ systems and apply it to the OMAS to show that ultimately bounded consensus tracking can be achieved if the network switching satisfies the piecewise average dwell time and activation time ratio conditions. Particularly, for vanishing perturbations, asymptotic tracking can be ensured under weaker switching conditions.
\end{abstract}

\begin{keyword}
 Open multi-agent systems, switched systems, repelling interactions, perturbations.
\end{keyword}

\end{frontmatter}

\section{Introduction}\label{sec:introduction}
Open multi-agent systems (OMASs) extend traditional multi-agent systems by allowing agents to dynamically join or leave the group \cite{hendrickx2016open}. Such migrations of agents create a naturally switching and size-varying network structure. By virtue of this feature, the OMAS has shown superiority over its traditional counterpart in covering a wide range of practical networks with variable structures like mobile ad-hoc networks, microgrids with plug-and-play control, and biomes with migrating species. On the other hand, the mediating effect of agent migrations makes an OMAS more prone to environmental perturbations, such as the lane change maneuver of joining vehicles that bring abrupt disturbances to the network dynamics/state of a vehicle platoon \cite{RN4530}. Meanwhile, agent migrations can also affect the normal cooperative interactions among agents, turning them into or directly bringing in antagonistic interactions. For example, a cyber-physical  system can be vulnerable to trojan attacks which connect disguised malicious devices to the network to convert normal connections into or create sabotage connections \cite{RN8295}. OMASs suffering these migration-induced adversities can be constantly perturbed or exhibit instability which impede desired coordination goals from being achieved. Given these facts, it is thus of both theoretical and practical significance to focus on the migration-induced adversities against the OMAS.

OMASs have become one of the emerging topics in the control community over the last few years sparked by \cite{hendrickx2016open,RN3365}. Two notable concepts of ``scale-independent quantities'' \cite{hendrickx2016open} and ``open distance'' \cite{RN3365} were developed to handle the size-varying nature of OMAS networks. Several subsequent results following these lines have been reported recently \cite{RN8530,RN5328,11345953,RN8526}.
Further, \cite{S2} provided an alternative line featuring the concept of multi-mode multi-dimensional ($M^3D$) systems \cite{verriest2012pseudo} to integrate the migration-induced  dimension variations and non-vanishing impulsive perturbations into a unified framework. Note that \cite{S2} studied only the leaderless consensus problem under cooperative interactions, which may not well reflect some practical scenarios such as disaster rescues that entail guidance and the cooperative interactions may be altered by the adverse environment. More recently, a first exploration on OMASs with antagonistic interactions has been made by
 \cite{11312939}, where the bipartite consensus problem was studied for OMASs under opposing interactions \cite{RN3154}. However, the results on OMASs with the more practical repelling interactions \cite{RN3154}, which can lead to Laplacians comprising eigenvalues with negative real parts, are still limited. Besides, most existing results only apply to OMASs without non-vanishing dynamics/state perturbations \cite{RN8530,RN9147}, or those with undirected/connected graphs \cite{RN5328,RN8526}, which also restrict the application scenarios.

Motivated by the above, in this work we focus on the consensus tracking problem of OMASs with non-vanishing perturbations and repelling antagonistic interactions.
The main contributions are as follows:
\begin{enumerate}
\item The consensus tracking problem of OMASs with migration-induced non-vanishing dynamics/state perturbations and antagonistic interactions is addressed, which extends the result of \cite{S2} that considered leaderless consensus for OMASs under only state perturbations and cooperative interactions.
\item The consideration of repelling interactions compensates the results for OMASs with purely cooperative interactions \cite{RN8530,RN8526} or opposing antagonistic interactions \cite{11312939}, and it shows that the OMAS must exhibit instability regardless of the network  connectivity if the repelling interactions dominate the cooperative ones.
\item A new $M^3D$ system-based analysis method, which accommodates non-vanishing perturbations and destabilizing repelling interactions while no longer requires the ``alternating switching'' condition as in \cite{S2}, is proposed for the OMAS over the signed graph that is allowed to be directed and intermittently disconnected compared to \cite{RN5328,RN8526}.
\end{enumerate}

The rest of this work is organized as follows. The system formulation and the related concepts are given in Section \ref{Sec_2} as preliminaries. The main results including the consensus tracking analysis for the OMAS are presented in Section \ref{Sec_3}. Section \ref{Sec_4_C} provides a simulation example for verifying the obtained results. Section \ref{Sec_5} concludes the work.

\textit{Notations}: $\textbf{1}_n$ denotes an $n\times 1$ vector of ones and $\textbf{0}_{m\times n}$ (resp. $\textbf{0}_n$) denotes an $m\times n$ (resp. $n\times 1$) matrix of zeros;
$\mathbb{N}$ and $\mathbb{N}_{>0}$ denote the sets of nonnegative and positive integers, respectively; $\mathbb{R}^{m\times n}$ (resp. $\mathbb{R}^{n}$) denotes the set of $m\times n$ (resp. $n\times 1$) real matrices and $\mathbb{R}_{\geq 0}$ denotes the set of nonnegative real  numbers;
$\mathbb{B}^{m\times n}$ represents the set of $m\times n$ binary (0-1) matrices; $|\mathcal{S}|$ denotes the cardinality of the set $\mathcal{S}$; $\mathrm{Re}(\cdot)$ denotes the real part of a complex number;
$\lambda(F)$ denotes the spectrum of any square matrix $F$, $\lambda_i(F)$ is the $i$-th eigenvalue of $F$, $\alpha(F)=\max_{i}\mathrm{Re}(\lambda_i(F))$ is the spectral abscissa of $F$, $\mathrm{trace}(F)$ denotes the trace of $F$; $\|\cdot\|$ denotes the 2-norm of a matrix or vector; for a real symmetric matrix $P$, $\lambda_{\max}(P)$ (resp. $\lambda_{\min}(P)$) denotes its maximum (resp. minimum) eigenvalue, $P \leq 0$ (resp. $P>0$) means negative semi-definite (resp. positive definite);
  $\mathrm{diag}(A_1,A_2,...,A_N)$ denotes a block diagonal matrix composed of matrices or scalars  $A_1,A_2,...,A_N$.

\section{Preliminaries}\label{Sec_2}
In this section, the system formulation and related concepts are presented. Note that in this work we consider any time-dependent function on a generic time interval $[t_0,t_f]$, $0\leq t_0<t_f <  +\infty$. When $t_f\rightarrow+\infty$, the interval becomes $[t_0,+\infty)$.

\subsection{System formulation}\label{Sec_2_A}
\noindent \textbf{1) Topology setting:}
Consider an OMAS network containing $N_{\sigma(t)}$ agents and a reference/leader agent. Suppose the reference does not receive information from any agent at any time. The OMAS topology is then represented by a switching digraph  $\tilde{\mathcal{G}}_{\sigma(t)}=(\tilde{\mathcal{V}}_{\sigma(t)},\tilde{\mathcal{E}}_{\sigma(t)})$, where ${\sigma}:\mathbb{R}_{\geq 0}\rightarrow {\mathcal{P}}$ is a c\`{a}dl\`{a}g piecewise constant function denoting the switching signal with $\mathcal{P}$ a finite set containing the indices of switching modes of ${\sigma}(t)$; $\tilde{\mathcal{V}}_{\sigma(t)}=\mathcal{V}_{\sigma(t)}\cup\{0\}$ where ${\mathcal{V}}_{\sigma(t)}=\{1,...,N_{\sigma(t)}\}$ denotes the index set of vertices/agents and $0$ is the index of the reference agent; $\tilde{\mathcal{E}}_{\sigma(t)}=\{(0,j)\}\cup\mathcal{E}_{\sigma(t)}$ where $\mathcal{E}_{{\sigma}(t)}\subseteq\mathcal{V}_{\sigma(t)}\times {\mathcal{V}}_{\sigma(t)}$ denotes the edge set in which for any $i,j\in\mathcal{V}_{\sigma(t)}$, $(j,i)$ denotes a directed edge from $j$ to $i$, $(0,j)$ denotes a directed edge from $0$ to $j$, $j\in\{i\in\mathcal{E}_{\sigma(t)} \mid d_i(\sigma(t))\neq 0\}$,  $d_{i}(\sigma(t))$ is an indicator of the presence of a directed edge from $0$ to $i$ at $t$, $d_i(\sigma(t))=1$ (resp. $d_i(\sigma(t))=-1$) if such an edge exists and is positive (resp. negative), $d_i(\sigma(t))=0$ if there exists no such an edge. Assume that for each $i\in\tilde{\mathcal{V}}_{\sigma(t)}$, $(i,i)\not\in \tilde{\mathcal{E}}_{{\sigma}(t)}$ at any $t$, i.e., no self-loop exists.
Let $\mathcal{E}^+_{{\sigma}(t)}$ and $\mathcal{E}^-_{{\sigma}(t)}=\mathcal{E}_{{\sigma}(t)}\setminus\mathcal{E}^+_{{\sigma}(t)}$ respectively be the sets of all the positive and negative edges in  $\mathcal{E}_{{\sigma}(t)}$. Let $[{a}_{ij}({{\sigma}}(t))]\in\mathbb{R}^{{N}_{\sigma(t)}\times {N}_{\sigma(t)}}$ denote the adjacency matrix of ${\mathcal{G}}_{{\sigma}(t)}=(\mathcal{V}_{\sigma(t)},\mathcal{E}_{\sigma(t)})$, where ${a}_{ij}({\sigma}(t))=0$ if $(j,i)\not\in{\mathcal{E}}_{{\sigma}(t)}$, ${a}_{ij}(\sigma(t))=1$ if $(j,i)\in{\mathcal{E}}^+_{{\sigma}(t)}$, ${a}_{ij}(\sigma(t))=-1$ if $(j,i)\in{\mathcal{E}}^-_{{\sigma}(t)}$. Then, define ${L}_{{\sigma}(t)}=[{l}_{ij}({\sigma}(t))]$, where $l_{ij}({\sigma}(t))=-{a}_{ij}({\sigma}(t))$ if $i\neq j$ and
 ${l}_{ij}({\sigma}(t))=\sum^{N_{\sigma(t)}}_{j=1}{a}_{ij}({\sigma}(t))$ if $i=j$, as the \textit{repelling} Laplacian matrix
\cite{RN3154} of ${\mathcal{G}}_{{\sigma}(t)}$ at any $t$.  Accordingly, the repelling Laplacian of $\tilde{\mathcal{G}}_{\sigma(t)}$ is defined as  $\tilde{{L}}_{{\sigma}(t)}=[\tilde{l}_{ij}(\sigma(t))]=[0,\textbf{0}^{\top}_{N_{\sigma(t)}};-\tilde{d}_{\sigma(t)},{Z}_{{\sigma}(t)}]\in\mathbb{R}^{(N_{\sigma(t)}+1)\times (N_{\sigma(t)}+1)}$, $\forall i,j\in\tilde{\mathcal{V}}_{\sigma(t)}$, where
 $\tilde{d}_{\sigma(t)}=[d_{1}({\sigma}(t)),...,d_{N_{\sigma(t)}}({\sigma}(t))]^{\top}$, ${Z}_{{\sigma}(t)}=[z_{ij}({\sigma}(t))]=L_{\sigma(t)}+D_{{\sigma}(t)}\in\mathbb{R}^{N_{\sigma(t)}\times N_{\sigma(t)}}$, $\forall i,j\in{\mathcal{V}}_{\sigma(t)}$,  $D_{\sigma(t)}=\mathrm{diag}(d_{1}({\sigma}(t)),...,d_{N_{\sigma(t)}}({\sigma}(t)))$. By definition, any repelling Laplacian matrix  has zero as one of its eigenvalues and can have eigenvalues with negative real parts.

Moreover, let $\tilde{\mathcal{E}}^+_{{\sigma}(t)}$ and $\tilde{\mathcal{E}}^-_{{\sigma}(t)}=\tilde{\mathcal{E}}_{{\sigma}(t)}\setminus\tilde{\mathcal{E}}^+_{{\sigma}(t)}$ be respectively the sets of all the positive and negative edges in $\tilde{\mathcal{E}}_{{\sigma}(t)}$. Then, let
 $\mathcal{P}^+=\{\phi\in\mathcal{P} \mid \tilde{\mathcal{E}}^-_{\phi}=\emptyset\}$ be the set of mode indices under which each $\tilde{\mathcal{G}}_{\phi}$ has only positive edges, $\mathcal{P}^+_{s}=\{\phi\in\mathcal{P}^+ \mid \tilde{\mathcal{G}}_{\phi} \text{ has a directed spanning tree rooted at 0}\}$, and  $\mathcal{P}^+_u=\mathcal{P}^+\setminus \mathcal{P}^+_s$; let $\mathcal{P}^-_{\exists}=\{\phi\in\mathcal{P} \mid \tilde{\mathcal{E}}^-_{\phi}\neq\emptyset\}=\mathcal{P}\setminus\mathcal{P}^+$ be the set of mode indices under which each $\tilde{\mathcal{G}}_{\phi}$ has at least one negative edge; let $\mathcal{P}^-_{>}=\{\phi\in\mathcal{P}^-_{\exists} \mid |\tilde{\mathcal{E}}^-_{\phi}|>|\tilde{\mathcal{E}}^+_{\phi}|\}=\{\phi\in\mathcal{P}^-_{\exists} \mid \sum_{i\in\tilde{\mathcal{V}}_{\phi}}\sum_{j\in\tilde{\mathcal{V}}_{\phi}\setminus\{i\}}\tilde{l}_{ij}(\phi)>0\}$ be the set of mode indices under which each $\tilde{\mathcal{G}}_{\phi}$ has more negative edges than positive edges.
 \begin{remark}
    It is clear that $\tilde{L}_{\phi}$ and $Z_{\phi}$ are both $M$-matrices for any $\phi\in\mathcal{P}^+$. This is, however, not necessarily the case for any $\phi\in\mathcal{P}^-_{\exists}$ due to the presence of negative edges.
 \end{remark}

The following assumptions are made for the digraph $\tilde{\mathcal{G}}_{\sigma(t)}$.
\begin{assumption}\label{connected_top}
   $\mathcal{P}^+_s\neq \emptyset$ and $\mathcal{P}^-_{\exists}\neq \emptyset$.
\end{assumption}
\begin{remark}\label{remark_Z}
  Assumption \ref{connected_top} indicates at least one digraph with only positive edges has a directed spanning tree rooted at 0 and rules out the trivial case of no digraph with negative edges. Meanwhile, 1) for any $\phi\in\mathcal{P}^+_s$, each eigenvalue of $Z_{\phi}$ has a positive real part; 2) for any $\phi\in\mathcal{P}^+_u$, at least one eigenvalue of $Z_{\phi}$ is zero \cite{S4}.
\end{remark}

\begin{assumption}\label{top_assump}
$\mathcal{P}^-_{\exists}=\mathcal{P}^-_{>}$.
\end{assumption}
\begin{remark}
Assumption \ref{top_assump} indicates that any digraph containing negative edges has more negative edges than positive edges, which implies $\mathcal{P}^+\cup\mathcal{P}^-_{>}=\mathcal{P}$. Such an assumption can reflect some practical scenarios. For example, DDoS attacks usually inject a large number of malicious connections to a cyber-physical system to overwhelm the normal connections and to further exhaust the resource while cause instability of the network \cite{RN3717}.
\end{remark}

\noindent\textbf{2) Agent dynamics:}
Denote the $k$-th switching instant of $\sigma(t)$ by $t_k$, $k\in\mathbb{N}_{>0}$. For each agent $i\in\mathcal{V}_{\sigma(t)}$ at any $t\in[t_k,t_{k+1})$, $k\in\mathbb{N}$, consider the following perturbed linear dynamics with a classic consensus tracking protocol:
\begin{align}\label{OMAS}
\dot{\xi}_{i}(t)=&A\xi_{i}(t)+\varrho\Big(\sum\limits_{j=1}^{N_{\sigma(t)}}a_{ij}(\sigma(t))({\xi}_{i}(t)-{\xi}_{j}(t))\nonumber\\&+d_i(\sigma(t))({\xi}_{i}(t)-\xi_0(t))\Big)+h_i(t),
\end{align}
where $\xi_{i}(t)\in\mathbb{R}^p$ is the state of agent $i$,
$A\in\mathbb{R}^{p\times p}$ and $\alpha(A)\geq 0$; $h_i(t)\in\mathbb{R}^p$ is the dynamics perturbation of $i$; $\varrho$ is a constant scalar to be determined, $\xi_0(t)\in\mathbb{R}^p$ is the state of the reference agent 0:
\begin{align}\label{exo_system}
\dot{\xi}_0(t)=A\xi_0(t),
\end{align}
where agent 0 has the same open-loop dynamics as each agent $i\in\mathcal{V}_{\sigma(t)}$. Further, the following assumption is made on the agents.
\begin{assumption}\label{exo_system_assump}
Any agent $i\in\mathcal{V}_{\sigma(t)}$ can migrate at $t_k$, $k\in\mathbb{N}_{>0}$ while  agent 0 does not migrate and suffers no state jump at any time.
\end{assumption}

Denote the stacked state of the agents $i\in\tilde{\mathcal{V}}_{\sigma(t)}$ by $\tilde{\xi}_{\sigma(t)}(t)=[\xi^{\top}_0(t), \xi^{\top}_{1}(t),...,\xi^{\top}_{N_{\sigma(t)}}(t)]^{\top}\in\mathbb{R}^{p(N_{\sigma(t)}+1)}$. Then, the compact form of \eqref{OMAS} and \eqref{exo_system} can be formulated as:
\begin{align}\label{OMAS_compact}
 \dot{\tilde{\xi}}_{\sigma(t)}(t)=(I_{N_{\sigma(t)}+1}\otimes  A+ \varrho \tilde{L}_{\sigma(t)}\otimes I_p) \tilde{\xi}_{\sigma(t)}(t)+\widehat{\tilde{h}}_{\sigma(t)}(t),
\end{align}
in which $\widehat{\tilde{h}}_{\sigma(t)}(t)=[\textbf{0}_{p};\tilde{h}_{\sigma(t)}(t)]\in\mathbb{R}^{p(N_{\sigma(t)}+1)}$ is the dynamics perturbation of the OMAS with $\tilde{h}_{\sigma(t)}(t)=[h^{\top}_1(t),...,h^{\top}_{N_{\sigma(t)}}(t)]^{\top}$ whose value can be unknown but has a finite bound $\overline{h}=\sup_{t\in[t_0,t_f]}\|\tilde{h}_{\sigma(t)}(t)\|\geq 0$.
The state transition of the OMAS at each $t_k$, $k\in\mathbb{N}_{>0}$  is given by:
\begin{align}\label{Agent_moves}
\tilde{\xi}_{\sigma(t_k^+)}(t_{k}^+)=&\widehat{\tilde{\Xi}}_{\sigma(t_k^+),\sigma(t_k^-)}\tilde{\xi}_{\sigma(t^-_k)}(t_{k}^-)+\widehat{\tilde{\Phi}}_k,
\end{align}
where the matrix
$\widehat{\tilde{\Xi}}_{\sigma(t_k^+),\sigma(t_k^-)}=[I_p,\textbf{0}^{\top}_{N_{\sigma(t_k^-)}}\otimes I_p;\textbf{0}_{N_{\sigma(t_k^+)}}\otimes I_p,\tilde{\Xi}_{\sigma(t_k^+),\sigma(t_k^-)}]\in\mathbb{B}^{p(N_{\sigma(t_k^+)}+1)\times p(N_{\sigma(t_k^-)}+1)}$ denotes the pure size variation of the OMAS network $\tilde{\mathcal{G}}_{\sigma(t)}$ under certain agent migrations at $t_k$ and the vector $\widehat{\tilde{\Phi}}_k=[\textbf{0}_{p};\widehat{\Phi}_{k}]\in\mathbb{R}^{p(N_{\sigma(t_k^+)}+1)}$ denotes the state jump of the OMAS at $t_k$ that is not brought by a pure size variation, where $\widehat{\Phi}_{k}\in\mathbb{R}^{pN_{\sigma(t_k^+)}}$ will be defined later. Specifically, $\tilde{\Xi}_{\sigma(t_k^+),\sigma(t_k^-)}=\Xi_{\sigma(t_k^+),\sigma(t_k^-)}\otimes I_p$ where $\Xi_{\sigma(t_k^+),\sigma(t_k^-)}\in\mathbb{B}^{N_{\sigma(t_k^+)}\times N_{\sigma(t_k^-)}}$ is obtained by:
\begin{itemize}
  \item inserting $s$ row vectors of zeros into $I_{N_{\sigma(t_k^-)}}$ as the new $i_1$-th,...,$i_s$-th rows if $s$ new agents join the OMAS at $t_k$ as the agents $i_1$,...,$i_s$ in $\mathcal{V}_{\sigma(t^+_k)}$;
  \item deleting the $i_1$-th,...,$i_s$-th rows from $I_{N_{\sigma(t_k^-)}}$ if the agents $i_1$,...,$i_s$ in $\mathcal{V}_{\sigma(t^-_k)}$ leave the OMAS at $t_k$;
  \item setting the value to $I_{N_{\sigma(t_k^-)}}$ if no agent migrates at $t_k$.
\end{itemize}
The vector
$\widehat{\Phi}_k=\widehat{\Phi}^{ind}_k+\widehat{\Phi}^{dep}_k$ denotes the state jump of the system \eqref{OMAS} at $t_k$ that is not brought by a pure size variation of ${\mathcal{G}}_{\sigma(t)}$, where $\widehat{\Phi}^{ind}_k$ is a \textit{state-independent} jump/impulse whose value does not depend on the OMAS state $\tilde{\xi}_{\sigma(t)}(t)$ for any $k\in\mathbb{N}_{>0}$ but has a finite bound $\overline{\Phi}=\sup_{k\in\mathbb{N}_{>0}}\|\widehat{\Phi}^{ind}_k\|\in[0,+\infty)$, $\widehat{\Phi}^{dep}_k={\breve{\Xi}}_{\sigma(t_k^+),\sigma(t_k^-)}\tilde{\xi}_{\sigma(t_k^-)}(t_k^-)$ is a \textit{state-dependent} impulse w.r.t. the OMAS state $\tilde{\xi}_{\sigma(t)}(t)$, where ${\breve{\Xi}}_{\sigma(t_k^+),\sigma(t_k^-)}\in\mathbb{R}^{pN_{\sigma(t_k^+)}\times p(N_{\sigma(t_k^-)}+1)}$ is to be determined.
Set ${\breve{\Xi}}_{\sigma(t_k^+),\sigma(t_k^-)}=\widehat{\Xi}_{\sigma(t_k^+),\sigma(t_k^-)}\tilde{\Upsilon}_{\sigma(t_k^-)}-\widehat{\Upsilon}_{\sigma(t_k^+),\sigma(t_k^-)}+\tilde{\Xi}_{\sigma(t_k^+),\sigma(t_k^-)}\breve{\Upsilon}_{\sigma(t_k^-)}$, where $\widehat{\Xi}_{\sigma(t_k^+),\sigma(t_k^-)}\in\mathbb{R}^{pN_{\sigma(t_k^+)}\times pN_{\sigma(t_k^-)}}$ is given, $\tilde{\Upsilon}_{\sigma(t)}=[-\textbf{1}_{N_{\sigma(t)}},
  I_{N_{\sigma(t)}}]\otimes I_p$, $\widehat{\Upsilon}_{\sigma(t_k^+),\sigma(t_k^-)}=[-\textbf{1}_{N_{\sigma(t^+_k)}}, \textbf{0}_{N_{\sigma(t^+_k)}\times N_{\sigma(t_k^-)}}]\otimes I_p$, $\breve{\Upsilon}_{\sigma(t)}=[-\textbf{1}_{N_{\sigma(t)}}, \textbf{0}_{N_{\sigma(t)}\times N_{\sigma(t)}}]\otimes I_p$.
Clearly,  ${\widehat{\Phi}}_k$ is state-dependent if and only if $\widehat{\Phi}^{ind}_k\equiv 0$; otherwise ${\widehat{\Phi}}_k$ is state-independent.
\begin{remark}\label{remark_matrix}
A state-independent impulse is in general {non-vanishing} regardless of the evolution of $\tilde{\xi}_{\sigma(t)}(t)$. A state-dependent impulse is {vanishing} with a vanishing $\tilde{\xi}_{\sigma(t_k^-)}(t_k^-)$ as $k\rightarrow +\infty$.
\end{remark}

\subsection{Related concepts}
In this work, we are interested in the consensus tracking problem of the OMAS \eqref{OMAS_compact}-\eqref{Agent_moves}. Denote the state of the tracking error between each agent $i\in\mathcal{V}_{\sigma(t)}$ and agent 0 by $\varepsilon_{i}(t)=\xi_{i}(t)-\xi_0(t)$ and the stacked state by $\tilde{\varepsilon}_{\sigma(t)}(t)=[\varepsilon^{\top}_{1}(t),...,\varepsilon^{\top}_{N_{\sigma(t)}}(t)]^{\top}=\tilde{\Upsilon}_{\sigma(t)}\tilde{\xi}_{\sigma(t)}(t)=\hat{\tilde{\xi}}_{\sigma(t)}(t)-\textbf{1}_{N_{\sigma(t)}}\otimes \xi_0(t)\in\mathbb{R}^{pN_{\sigma(t)}}$, where $\hat{\tilde{\xi}}_{\sigma(t)}(t)=[\xi^{\top}_1(t),...,\xi^{\top}_{N_{\sigma(t)}}(t)]^{\top}\in\mathbb{R}^{pN_{\sigma(t)}}$. Then, the consensus tracking property of interest is defined as follows.
\begin{definition}\label{def_sys}
The group of agents \eqref{OMAS} is said to reach ultimately bounded consensus tracking of the reference agent \eqref{exo_system}, if there exists a class $\mathcal{KL}$ function $\beta$ and a certain constant $\epsilon\geq 0$ such that for any initial condition,
\begin{align}\label{Def_GUPS}
\|\tilde{\varepsilon}_{\sigma(t)}(t)\|\leq \beta(\|\tilde{\varepsilon}_{\sigma(t_0)}(t_0)\|,t-t_0)+\epsilon,\ \forall t\geq t_0,
\end{align}
holds. In particular, if \eqref{Def_GUPS} holds with $\epsilon=0$, then \eqref{OMAS} is said to reach (asymptotic) consensus tracking of \eqref{exo_system}.
\end{definition}

The following definition for the switching signal is needed:
\begin{definition}[\cite{S4}]\label{Def_Piecewise_ADT}
Denote the number of switchings of $\sigma(t)$ on $[t_k,t_f]\subseteq [t_0,t_f]$ by $\hat{N}(t_k,t_f)$ for each $k\in\mathbb{N}$. For the given constant $\hat{N}_0\geq 0$, the constant ${\tau}(t_k,t_f)$ satisfying
\begin{align}\label{Piecewise_ADT_slow}
\hat{N}(t_k,t_f)\leq \hat{N}_0+\frac{t_f-t_k}{{\tau}(t_k,t_f)},\ k\in\mathbb{N},
\end{align}
 is said to be the piecewise average dwell time (ADT) of the switching signal $\sigma(t)$ on $[t_k,t_f]$.
\end{definition}

\section{Main results}\label{Sec_3}
In this section, the main results including the consensus tracking analysis for the OMAS are presented.

Note that for the right-hand side of \eqref{Agent_moves}, one has: $\tilde{\Upsilon}_{\sigma(t_k^+)}\widehat{\tilde{\Xi}}_{\sigma(t_k^+),\sigma(t_k^-)}\tilde{\xi}_{\sigma(t_k^-)}(t_k^-)=\tilde{\Xi}_{\sigma(t_k^+),\sigma(t_k^-)}\hat{\tilde{\xi}}_{\sigma(t_k^-)}(t_k^-)-\textbf{1}_{N_{\sigma(t_k^+)}}\otimes \xi_0(t_k^-)$, $\widehat{\Upsilon}_{\sigma(t_k^+),\sigma(t_k^-)}\tilde{\xi}_{\sigma(t_k^-)}(t_k^-)=-\textbf{1}_{N_{\sigma(t_k^+)}}\otimes \xi_0(t_k^-)$, $\breve{\Upsilon}_{\sigma(t_k^-)}\tilde{\xi}_{\sigma(t_k^-)}(t_k^-)=-\textbf{1}_{N_{\sigma(t_k^-)}}\otimes \xi_0(t_k^-)$, $\tilde{\Upsilon}_{\sigma(t_k^+)}\widehat{\tilde{\Phi}}_k=\widehat{\Phi}_k$. Then, the tracking error system of $\tilde{\varepsilon}_{\sigma(t)}(t)$ is given by:
\begin{subnumcases}{\label{switched_sys111}}
\dot{\tilde{\varepsilon}}_{\sigma(t)}(t)=\tilde{A}_{\sigma(t)} \tilde{\varepsilon}_{\sigma(t)}(t)+\tilde{h}_{\sigma(t)}(t), \ t\in[t_{k-1},t_{k}),
\label{switched_sys_1}\\
\tilde{\varepsilon}_{\sigma(t_k^+)}(t_k^+)=\tilde{\Xi}_{\sigma(t_k^+),\sigma(t_k^-)}\tilde{\varepsilon}_{\sigma(t_k^-)}(t_k^-)+\tilde{\Phi}_k,\ k\in\mathbb{N}_{>0},
 \label{switched_sys_2}
 \end{subnumcases}
where
$\tilde{A}_{\sigma(t)}= I_{N_{\sigma(t)}}\otimes A+\varrho Z_{\sigma(t)} \otimes I_p\in\mathbb{R}^{pN_{\sigma(t)}\times pN_{\sigma(t)}}$; the perturbation $\tilde{h}_{\sigma(t)}(t)$ is non-vanishing if
$\|\tilde{h}_{\sigma(t)}(t)\|\not\rightarrow 0$ as $\|\tilde{\varepsilon}_{\sigma(t)}(t)\|\rightarrow 0$; $\tilde{\Phi}_k=\tilde{\Phi}^{ind}_k+\tilde{\Phi}^{dep}_k$ in which $\tilde{\Phi}^{ind}_k=\widehat{\Phi}_k^{ind}$ is a state-independent impulse and  $\tilde{\Phi}^{dep}_k=\widehat{\Xi}_{\sigma(t_k^+),\sigma(t_k^-)}\tilde{\varepsilon}_{\sigma(t_k^-)}(t_k^-)$ is a state-dependent impulse, both w.r.t. the error state $\tilde{\varepsilon}_{\sigma(t)}(t)$.  It can be seen that $\tilde{\Phi}_k$ is state-dependent  (resp. state-independent) if and only if $\widehat{\Phi}^{ind}_k\equiv 0$ (resp. $\widehat{\Phi}^{ind}_k\not\equiv 0$).
Moreover, \eqref{switched_sys_2} can be rewritten by merging $\tilde{\Phi}_k^{dep}$ into the state-dependent term:
\begin{align}\label{M3D_vanishing}
\tilde{\varepsilon}_{\sigma(t_k^+)}(t_k^+)=\breve{\tilde{\Xi}}_{\sigma(t_k^+),\sigma(t_k^-)}\tilde{\varepsilon}_{\sigma(t_k^-)}(t_k^-)+\tilde{\Phi}_k^{ind},\ k\in\mathbb{N}_{>0},
\end{align}
where $\breve{\tilde{\Xi}}_{\sigma(t_k^+),\sigma(t_k^-)}=\tilde{\Xi}_{\sigma(t_k^+),\sigma(t_k^-)}+\widehat{\Xi}_{\sigma(t_k^+),\sigma(t_k^-)}$.
\begin{remark}
It is clear to see that both the OMAS \eqref{OMAS_compact}-\eqref{Agent_moves} and the tracking error system \eqref{switched_sys111} are $M^3D$ systems featuring different subsystem dimensions as in \cite{S2}, except we further consider the non-vanishing perturbation $\tilde{h}_{\sigma(t)}(t)$. The non-vanishing property of the perturbation is similar in definition to that in \cite{khalil2002noninear}.
\end{remark}
The following lemma is for the matrices $\tilde{L}_{\sigma(t)}$ and $Z_{\sigma(t)}$:
\begin{lemma}\label{Z_eig}
For any $\phi\in\mathcal{P}^-_{>}$, each of $\tilde{L}_{\phi}$ and $Z_{\phi}$ has at least one eigenvalue with a negative real part.
\end{lemma}
\begin{pf}
   For any $\phi\in\mathcal{P}$, recall that $\tilde{{L}}_{\phi}=[\tilde{l}_{ij}(\phi)]=[0,\textbf{0}^{\top}_{N_{\phi}};-\tilde{d}_{\phi},{Z}_{\phi}]$ where $\tilde{d}_{\phi}=[d_{1}(\phi),...,d_{N_{\phi}}(\phi)]^{\top}$, $Z_{\phi}=L_{\phi}+D_{\phi}$ with $D_{\phi}=\mathrm{diag}(d_{1}(\phi),...,d_{N_{\phi}}(\phi))$, $\tilde{l}_{ii}(\phi)=-\sum_{j\neq i}\tilde{l}_{ij}(\phi)$ for any $i\in\tilde{\mathcal{V}}_{\phi}$, $\tilde{l}_{ij}(\phi)=l_{ij}(\phi)=-a_{ij}(\phi)$ for any $i,j\in{\mathcal{V}}_{\phi}$, $i\neq j$, $\tilde{l}_{i0}(\phi)=-d_i(\phi)$ for any $i\in\mathcal{V}_{\phi}$, and $\tilde{l}_{0j}(\phi)=0$ for any $j\in\tilde{\mathcal{V}}_{\phi}$. Since $\sum_{i\in\tilde{\mathcal{V}}_{\phi}}\sum_{j\in\tilde{\mathcal{V}}_{\phi}\setminus\{i\}}\tilde{l}_{ij}(\phi)>0$ for any $\phi\in\mathcal{P}^-_{>}$, then it follows that $\sum_{i}\lambda_i(\tilde{L}_{\phi})=\mathrm{trace}(\tilde{L}_{\phi})=\sum_{i\in\tilde{\mathcal{V}}_{\phi}}\tilde{l}_{ii}(\phi)=-\sum_{i\in\tilde{\mathcal{V}}_{\phi}}\sum_{j\in\tilde{\mathcal{V}}_{\phi}\setminus\{i\}}\tilde{l}_{ij}(\phi)<0$, i.e., $\tilde{L}_{\phi}$ has at least one eigenvalue with a negative real part. For any $\phi\in\mathcal{P}^-_{>}$ one has $\sum_{i\in\tilde{\mathcal{V}}_{\phi}}\sum_{j\in\tilde{\mathcal{V}}_{\phi}\setminus\{i\}}\tilde{l}_{ij}(\phi)=\sum_{i\in{\mathcal{V}}_{\phi}}\sum_{j\in{\mathcal{V}}_{\phi}\setminus\{i\}}{l}_{ij}(\phi)-\sum_{i\in{\mathcal{V}}_{\phi}}d_i(\phi)=-\sum_{i\in{\mathcal{V}}_{\phi}}{l}_{ii}(\phi)-\sum_{i\in{\mathcal{V}}_{\phi}}d_i(\phi)>0$, which implies $\sum_{i}\lambda_i(Z_{\phi})=\mathrm{trace}({Z}_{\phi})=\sum_{i\in\mathcal{V}_{\phi}}({l}_{ii}(\phi)+d_{i}(\phi))<0$, i.e., $Z_{\phi}$ has at least one eigenvalue with a negative real part. $\blacksquare$
  \end{pf}
\begin{remark}
 Lemma \ref{Z_eig} implies each of $\tilde{L}_{\phi}$ and $Z_{\phi}$ has at least one eigenvalue with a negative real part regardless of the connectivity of $\tilde{\mathcal{G}}_{\phi}$ as long as $\phi\in\mathcal{P}^-_{>0}$. Such a conclusion may not be true if the digraph has negative edges no more than positive edges.
 Note that Lemma \ref{Z_eig} can be readily applied to the more general case of weighted edges since $\phi\in\mathcal{P}^-_{>0}$ also indicates $\sum_{i\in\tilde{\mathcal{V}}_{\phi}}\sum_{j\in\tilde{\mathcal{V}}_{\phi}\setminus\{i\}}\tilde{l}_{ij}(\phi)>0$.
\end{remark}

The next lemma on the Kronecker sum is needed for the upcoming analysis.
\begin{lemma}[\cite{bernstein2009matrix}]\label{MAS_tra1}
Given $G\in\mathbb{R}^{n\times n}$ and $H\in\mathbb{R}^{r\times r}$, let $Y= G\otimes I_r+I_n\otimes H$. Then, $\lambda(Y)=\{\lambda_G+\lambda_H \mid \lambda_G\in\lambda(G),\lambda_H\in\lambda(H)\}$.
\end{lemma}

Moreover, set $\varrho<\min_{\phi\in\mathcal{P}^+_s}\frac{\alpha({A}_{\phi})}{\alpha(-Z_{\phi})}$ in \eqref{switched_sys111}. Then, we have the following lemma.

\begin{lemma}\label{abs_A}
Consider \eqref{switched_sys111} with $\varrho<\min_{\phi\in\mathcal{P}^+_s}\frac{\alpha({A}_{\phi})}{\alpha(-Z_{\phi})}$ under Assumption \ref{connected_top}. Then, $\alpha(\tilde{A}_{\phi})<0$ for any $\phi\in\mathcal{P}^+_s$, $\alpha(\tilde{A}_{\phi})\geq 0$ for any $\phi\in\mathcal{P}^+_{u}$, and $\alpha(\tilde{A}_{\phi})> 0$ for any $\phi\in\mathcal{P}^-_{>}$.
\end{lemma}
\begin{pf}
Given any $\phi\in\mathcal{P}$, recall that $\alpha(A_{\phi}) \geq 0$. Since $\alpha(-Z_{\phi})<0$ for any $\phi\in\mathcal{P}^+_s$ under Assumption \ref{connected_top}, and $\varrho<\min_{\phi\in\mathcal{P}^+_s}\frac{\alpha({A}_{\phi})}{\alpha(-Z_{\phi})}\leq 0$, it can then be obtained from Lemma \ref{MAS_tra1} that $\alpha(\tilde{A}_{\phi})=\alpha({A}_{\phi})+\alpha(\varrho Z_{\phi})<0$ for any $\phi\in\mathcal{P}^+_s$.  Similarly, since at least one eigenvalue of $Z_{\phi}$ is zero for any $\phi\in\mathcal{P}^+_{u}$ (Remark \ref{remark_Z}) and at least one eigenvalue of $Z_{\phi}$ has a negative real part  for any $\phi\in\mathcal{P}^-_{>}$ (Lemma \ref{Z_eig}), then by Lemma \ref{MAS_tra1} one obtains  that at least one eigenvalue of $\tilde{A}_{\phi}$ has a nonnegative real part, i.e., $\alpha(\tilde{A}_{\phi})\geq 0$, for any $\phi\in\mathcal{P}^+_{u}$, and at least one eigenvalue of $\tilde{A}_{\phi}$ has a positive real part, i.e., $\alpha(\tilde{A}_{\phi})>0$, for any $\phi\in\mathcal{P}^-_{>}$, respectively. $\blacksquare$
\end{pf}
\begin{remark}
Lemma \ref{abs_A} implies that the nominal subsystem $\dot{\tilde{\varepsilon}}_{\phi}(t)=\tilde{A}_{\phi} \tilde{\varepsilon}_{\phi}(t)$ of \eqref{switched_sys_1} is: 1) asymptotically stable if $\tilde{\mathcal{G}}_{\phi}$ with only positive edges has a directed spanning tree rooted at agent 0 ($\phi\in\mathcal{P}^+_s$); 2)  not asymptotically stable (marginally stable or unstable) if $\tilde{\mathcal{G}}_{\phi}$ with only positive edges has no directed spanning tree or $\tilde{\mathcal{G}}_{\phi}$ has more negative edges than positive edges ($\phi\in\mathcal{P}^+_u\cup\mathcal{P}^-_{>}$). In this sense, define $\mathcal{P}_s=\mathcal{P}^+_s$ and $\mathcal{P}_u=\mathcal{P}\setminus\mathcal{P}_s=\mathcal{P}^+_u\cup\mathcal{P}^-_{>}$.
\end{remark}

Based on the above, the consensus tracking analysis of the considered OMAS is presented as follows.

\begin{thm}\label{Thm_1}
Consider the OMAS \eqref{OMAS} and \eqref{exo_system} on $[t_0,t_f]$ with $\varrho<\min_{\phi\in\mathcal{P}^+_s}\frac{\alpha({A}_{\phi})}{\alpha(-Z_{\phi})}$ under Assumptions \ref{connected_top}, \ref{top_assump}, and \ref{exo_system_assump}.
The group of agents \eqref{OMAS} reaches ultimately bounded  consensus tracking of the reference agent \eqref{exo_system} under $\widehat{\Phi}^{ind}_k\not\equiv 0$ and non-vanishing $\tilde{h}_{\sigma(t)}(t)$, if $\sigma(t)$ satisfies
\begin{align}
\label{thm_ratio}
{T_{s}(t_j,t_f)}({\overline{\gamma}_s-\tilde{\gamma}})+&{T_{u}(t_j,t_f)}({\overline{\gamma}_u-\tilde{\gamma}}) \leq 0,\\
\label{thm_ADT}
{\tau}(t_j,t_f)\geq & -\frac{\ln\mu}{\tilde{\gamma}},
\end{align}
for any $j\in\{0,1,...,\hat{N}(t_0,t_f)\}$, where $T_s(a,b)$ and $T_u(a,b)$ respectively denote the total activation time of $\phi\in\mathcal{P}_s$ and  $\phi\in\mathcal{P}_u$ on the interval $[a,b]$, $\tilde{\gamma}\in(\overline{\gamma}_s,0)$, $\overline{\gamma}_s=\max_{\phi\in\mathcal{P}_s}\tilde{\gamma}_{\phi}$, $\tilde{\gamma}_{\phi}\in(\alpha(\tilde{A}_{\phi}),0)$ for any $\phi\in\mathcal{P}_s$, $\overline{\gamma}_u=\max_{\phi\in\mathcal{P}_u}\tilde{\gamma}_{\phi}$, $\tilde{\gamma}_{\phi}\in (\alpha(\tilde{A}_{\phi}),+\infty)$ for any $\phi\in\mathcal{P}_u$.
 The ultimate bound is given by $\epsilon=\sqrt{\underline{P}^{-1}}(\overline{c}(\frac{1-\mu^{\hat{N}_0+1}}{1-\mu}+\frac{e^{(1+\hat{N}_0)\ln\mu}}{1-\mathrm{e}^{\overline{\varsigma}}})+\overline{\Theta}(\frac{1-\mu^{\hat{N}_0}}{1-\mu}+\frac{e^{\hat{N}_0\ln\mu}}{1-\mathrm{e}^{\overline{\varsigma}}}))$, where $\overline{c}=\frac{\overline{\vartheta}}{-\tilde{\gamma}}$, $\overline{\varsigma}=\max_{j}(\tau(t_{j},t_f)\tilde{\gamma}+\ln\mu)$, $\overline{\vartheta}=\overline{h}\frac{\overline{P}}{\sqrt{\underline{P}}}$, $\underline{P}=\min_{\phi\in\mathcal{P}}\lambda_{\min}(\tilde{P}_{\phi})$ and $\overline{P}=\max_{\phi\in\mathcal{P}}\lambda_{\max}(\tilde{P}_{\phi})$, $\tilde{P}_{\phi}$ is a positive definite matrix satisfying $\tilde{A}^{\top}_{\phi}\tilde{P}_{\phi}+\tilde{P}_{\phi}\tilde{A}_{\phi}\leq 2\tilde{\gamma}_{\phi}\tilde{P}_{\phi}$ for any $\phi\in\mathcal{P}$, $\mu=\sqrt{\frac{\overline{P}}{\underline{P}}}\max(1,\max_{\phi,\hat{\phi}}\|\breve{\tilde{\Xi}}_{\phi,\hat{\phi}}\|)>1$, $\overline{\Theta}=\overline{\Phi}\sqrt{\overline{P}}\geq 0$. Particularly, under $\widehat{\Phi}^{ind}_k\equiv 0$ and $\overline{h}=0$,  \eqref{OMAS} reaches (asymptotic) consensus tracking of \eqref{exo_system} if $\sigma(t)$ satisfies \eqref{thm_ratio} and \eqref{thm_ADT} for $j=0$.
\end{thm}
\begin{pf}
For any $t\in[t_k,t_{k+1})\subseteq [t_0,t_f]$, $k\in\mathbb{N}$, consider the function $W_{\sigma(t)}(t,\tilde{\varepsilon}_{\sigma(t)}(t))=\tilde{\varepsilon}^{\top}_{\sigma(t)}(t)\tilde{P}_{\sigma(t)}\tilde{\varepsilon}^{\top}_{\sigma(t)}(t)$ w.r.t. the tracking error system \eqref{switched_sys111}, where for each $\phi\in\mathcal{P}$, $\tilde{P}_{\phi}$ is a positive definite matrix satisfying $\tilde{A}^{\top}_{\phi}\tilde{P}_{\phi}+\tilde{P}_{\phi}\tilde{A}_{\phi}\leq 2\tilde{\gamma}_{\phi}\tilde{P}_{\phi}$, with $\tilde{\gamma}_{\phi}\in(\alpha(\tilde{A}_{\phi}),0)$ for any $\phi\in\mathcal{P}_s$ and $\tilde{\gamma}_{\phi}\in(\alpha(\tilde{A}_{\phi}),+\infty)$ for any $\phi\in\mathcal{P}_u$. Then, one obtains that $\underline{P}\|\tilde{\varepsilon}_{\sigma(t)}(t)\|^2\leq W_{\sigma(t)}(t,\tilde{\varepsilon}_{\sigma(t)}(t))\leq \overline{P}\|\tilde{\varepsilon}_{\sigma(t)}(t)\|^2$, where $\underline{P}=\min\limits_{\phi\in\mathcal{P}}\lambda_{\min}(\tilde{P}_{\phi})$ and $\overline{P}=\max\limits_{\phi\in\mathcal{P}}\lambda_{\max}(\tilde{P}_{\phi})$. Replacing $W_{\sigma(t)}(t,\tilde{\varepsilon}_{\sigma(t)}(t))$ with $W_{\sigma(t)}(t)$ for brevity, one derives:
\begin{align}\label{proof_V}
\dot{W}_{\sigma(t)}(t)\leq & \tilde{\varepsilon}^{\top}_{\sigma(t)}(t)(\tilde{A}^{\top}_{\sigma(t)}\tilde{P}_{\sigma(t)}+\tilde{P}_{\sigma(t)}\tilde{A}_{\sigma(t)})\tilde{\varepsilon}_{\sigma(t)}(t)\nonumber\\&+
 \tilde{h}^{\top}_{\sigma(t)}(t)\tilde{P}_{\sigma(t)}\tilde{\varepsilon}_{\sigma(t)}(t)+\tilde{\varepsilon}^{\top}_{\sigma(t)}(t)\tilde{P}_{\sigma(t)}\tilde{h}_{\sigma(t)}(t)\nonumber\\ \leq & \tilde{\varepsilon}^{\top}_{\sigma(t)}(t)(\tilde{A}^{\top}_{\sigma(t)}\tilde{P}_{\sigma(t)}+\tilde{P}_{\sigma(t)}\tilde{A}_{\sigma(t)})\tilde{\varepsilon}_{\sigma(t)}(t)\nonumber\\&+
 2\overline{h}\frac{\overline{P}}{\sqrt{\underline{P}}}\sqrt{W_{\sigma(t)}(t)}.
\end{align}
Moreover, it follows from \eqref{M3D_vanishing} that
\begin{align}\label{proof_jump}
\|\tilde{\varepsilon}_{\sigma(t_k^+)}(t_k^+)\|\leq & \|\breve{\tilde{\Xi}}_{\sigma(t_k^+),\sigma(t_k^-)}\|\|\tilde{\varepsilon}_{\sigma(t_k^-)}(t_k^-)\|+\|\tilde{\Phi}^{ind}_k\|,
\end{align}
which implies
\begin{align}\label{proof_jump_1}
\sqrt{\frac{1}{\overline{P}}W_{\sigma(t)}(t)} \leq & \|\breve{\tilde{\Xi}}_{\sigma(t_k^+),\sigma(t_k^-)}\|\sqrt{\frac{1}{\underline{P}}W_{\sigma(t)}(t)}+\overline{\Phi}.
\end{align}
Denoting $V_{\sigma(t)}(t,\tilde{\varepsilon}_{\sigma(t)}(t))=\sqrt{W_{\sigma(t)}(t,\tilde{\varepsilon}_{\sigma(t)}(t))}$ and using $V_{\sigma(t)}(t)$ instead of $V_{\sigma(t)}(t,\tilde{\varepsilon}_{\sigma(t)}(t))$ for brevity, one then obtains from the above that
\begin{align}
\label{thm_M3D_GUPS}
&\sqrt{\underline{P}}\|\tilde{\varepsilon}_{\sigma(t)}(t)\|\leq V_{\sigma(t)}(t)\leq \sqrt{\overline{P}}\|\tilde{\varepsilon}_{\sigma(t)}(t)\|,\\
\label{thm_M3D_GUPS_conv}
&\dot{V}_{\sigma(t)}(t)\leq \gamma_{\sigma(t)} V_{\sigma(t)}(t)+\overline{\vartheta},\\
\label{thm_M3D_GUPS_jump}
&V_{\sigma(t_k^+)}(t_k^+)\leq \mu V_{{\sigma}(t_k^-)}(t_k^-)+\overline{\Theta},
\end{align}
where the parameters $\gamma_{\sigma(t)}=\tilde{\gamma}_{\sigma(t)}$, $\overline{\vartheta}=\overline{h}\frac{\overline{P}}{\sqrt{\underline{P}}}$, $\mu=\sqrt{\frac{\overline{P}}{\underline{P}}}\max(1,\max\limits_{\phi,\hat{\phi}}\|\breve{\tilde{\Xi}}_{\phi,\hat{\phi}}\|)>1$, $\overline{\Theta}=\overline{\Phi}\sqrt{\overline{P}}$. Then, it follows from \eqref{thm_M3D_GUPS_conv} and \eqref{thm_M3D_GUPS_jump} that
  \begin{align}\label{proof_1}
V_{\sigma(t)}(t)
\leq & e^{{\gamma_{\sigma(t_k^+)}}(t-t_k)}(\mu V_{\sigma(t_k^-)}(t_k^-)+\overline{\Theta})\nonumber\\&+\int_{t^+_k}^{t} e^{{\gamma_{\sigma(t_k^+)}}(t-\tau)}{\overline{\vartheta}}d\tau \nonumber\\
\leq & e^{{\gamma_{\sigma(t_k^+)}}(t-t_k)}(\mu e^{{\gamma_{\sigma(t_{k-1}^+)}}(t_k-t_{k-1})}V_{\sigma(t_{k-1}^-)}(t_{k-1}^+)\nonumber\\&+\mu\int_{t^+_{k-1}}^{t_k^-} e^{{\gamma_{\sigma(t_{k-1}^+)}}(t_k-\tau)}{\overline{\vartheta}}d\tau+\overline{\Theta})\nonumber\\&+\int_{t^+_k}^{t} e^{{\gamma_{\sigma(t_k^+)}}(t-\tau)}{\overline{\vartheta}}d\tau\nonumber\\
\leq & \cdots \nonumber\\
\leq & \mu^k e^{\gamma_{\sigma(t_k^+)}(t-t_k)}e^{\sum\limits_{j=1}^{k}\gamma_{\sigma(t_{k-j}^+)}(t_{k-j+1}-t_{k-j})}\nonumber\\&\times V_{\sigma(t_{0}^+)}(t_{0}^+)+\int_{t^+_k}^{t} e^{{\gamma_{\sigma(t_k^+)}}(t-\tau)}{\overline{\vartheta}}d\tau\nonumber\\&+\sum\limits_{j=1}^{k}\mu^{j} e^{{\gamma_{\sigma(t_k^+)}}(t-t_k)+\sum\limits_{\nu=1}^{j-1}{\gamma_{\sigma(t_{k-\nu}^+)}(t_{k-\nu+1}-t_{k-\nu})}}\nonumber\\&\times\int_{t^+_{k-j}}^{t^-_{k-j+1}}e^{\gamma_{\sigma(t_{k-j}^+)}(t_{k-j+1}-\tau)}{\overline{\vartheta}}d\tau\nonumber\\&+\sum\limits_{j=1}^{k}\mu^{j-1}e^{{\gamma_{\sigma(t_k^+)}}(t-t_k)+\sum\limits_{\nu=1}^{j-1}\gamma_{\sigma(t_{k-\nu}^+)}(t_{k-\nu+1}-t_{k-\nu})}\nonumber\\&\times\overline{\Theta},
 \end{align}
  which, by \eqref{thm_ratio} and Definition \ref{Def_Piecewise_ADT}, further yields
 \begin{align}\label{proof_3}
V_{\sigma(t)}(t)
\leq &  e^{k\ln\mu+\overline{\gamma}_{s}T_s(t_0,t)+\overline{\gamma}_{u}T_u(t_0,t)}V_{\sigma(t_{0}^+)}(t_{0}^+)\nonumber\\&+\int_{t^+_k}^{t} e^{\hat{N}(\tau,t)\ln\mu+\overline{\gamma}_sT_s(\tau,t)+\overline{\gamma}_uT_u(\tau,t)}{\overline{\vartheta}}d\tau\nonumber\\&+\sum\limits_{j=1}^{k}\int_{t_{k-j}^+}^{t^-_{k-j+1}} e^{\hat{N}(\tau,t)\ln\mu+\overline{\gamma}_sT_s(\tau,t)+\overline{\gamma}_uT_u(\tau,t)}{\overline{\vartheta}}d\tau\nonumber\\&+\sum\limits_{j=1}^{k}\mu^{j-1}e^{\overline{\gamma}_sT_s(t_{k-j+1},t)+\overline{\gamma}_uT_u(t_{k-j+1},t)}\overline{\Theta}\nonumber\\
\leq & \underbrace{e^{\hat{N}_0\ln\mu+(\frac{\ln\mu}{\tau(t_0,t)}+\tilde{\gamma})(t-t_0)}V_{\sigma(t_{0}^+)}(t_{0}^+)}_{\tilde{\beta}(V_{\sigma(t_0)}(t_0),t-t_0)}\nonumber\\&+\underbrace{\overline{c}+\sum\limits_{j=1}^{k}\overline{c} e^{j\ln\mu+\tilde{\gamma}(t-t_{k-j+1})}}_{{\Omega}_k(t)}\nonumber\\&+\underbrace{\sum\limits_{j=1}^{k}e^{(j-1)\ln\mu+\tilde{\gamma}(t-t_{k-j+1})}\overline{\Theta}}_{\tilde{\Omega}_k(t)},
  \end{align}
  where $\overline{\gamma}_s=\max\limits_{\phi\in\mathcal{P}_s}{\gamma}_{\phi}$, $\overline{\gamma}_u=\max\limits_{\phi\in\mathcal{P}_u}{\gamma}_{\phi}$, and
  \begin{align}\label{proof_series}
\Omega_k(t) \leq & \sum\limits_{j=0}^{\hat{N}_0}\overline{c}e^{j\ln\mu}+\sum\limits_{j=\hat{N}_0+1}^{k}\overline{c}e^{j\ln\mu+\tilde{\gamma}\tau(t_{k-j+1},t)(j-1-\hat{N}_0)}\nonumber\\
\leq &
\underbrace{\overline{c}(\frac{1-\mu^{\hat{N}_0+1}}{1-\mu}+\hat{\alpha}\frac{1-\mathrm{e}^{\overline{\varsigma}(k-\hat{N}_0)}}{1-\mathrm{e}^{\overline{\varsigma}}})}_{\overline{\Omega}_k},
  \end{align}
   in which $\overline{c}=\frac{\overline{\vartheta}}{-\tilde{\gamma}}$, $\hat{\alpha}=e^{(1+\hat{N}_0)\ln\mu}$, $\overline{\varsigma}=\max\limits_{j}(\tau(t_{j},t_f)\tilde{\gamma}+\ln\mu)$, and
    \begin{align}\label{proof_series_1}
\tilde{\Omega}_k(t)
\leq &
\overline{\Theta}(\sum\limits_{j=1}^{\hat{N}_0}\mu^{j-1}+\hat{c}\sum\limits_{j=\hat{N}_0+1}^{k}e^{(j-1-\hat{N}_0)(\ln\mu+\tilde{\gamma}\tau(t_{k-j+1},t))})\nonumber\\
\leq &
\underbrace{\overline{\Theta}(\frac{1-\mu^{\hat{N}_0}}{1-\mu}+\hat{c}\frac{1-\mathrm{e}^{\overline{\varsigma}(k-\hat{N}_0)}}{1-\mathrm{e}^{\overline{\varsigma}}})}_{\overline{\tilde{\Omega}}_k},
  \end{align}
  where $\hat{c}=e^{\hat{N}_0\ln\mu}$. It can be seen that $\overline{\Omega}_k$ and $\overline{\tilde{\Omega}}_k$ are both monotonically increasing w.r.t. $k$.
  Note that by \eqref{thm_ADT}, it follows that $\overline{\varsigma}<0$ and $\tilde{\beta}$ is a class $\mathcal{KL}$ function, which implies $\limsup\limits_{k\rightarrow +\infty}\overline{\Omega}_k+\overline{\tilde{\Omega}}_k\leq \iota$, $\iota=\overline{c}(\frac{1-\mu^{\hat{N}_0+1}}{1-\mu}+\frac{\hat{\alpha}}{1-\mathrm{e}^{\overline{\varsigma}}})+\overline{\Theta}(\frac{1-\mu^{\hat{N}_0}}{1-\mu}+\frac{\hat{c}}{1-\mathrm{e}^{\overline{\varsigma}}})$. Meanwhile, one derives from \eqref{proof_3} that $V_{\sigma(t)}(t)\leq \tilde{\beta}(V_{\sigma(t_0)}(t_0),t-t_0)+\overline{\Omega}_k+\overline{\tilde{\Omega}}_k$, which by \eqref{thm_M3D_GUPS} implies $\|\tilde{\varepsilon}_{\sigma(t)}(t)\|\leq \beta(\|\tilde{\varepsilon}_{\sigma(t_0)}(t_0)\|,t-t_0)+\sqrt{\underline{P}^{-1}}(\overline{\Omega}_k+\overline{\tilde{\Omega}}_k)$, where $\beta(\|\tilde{\varepsilon}_{\sigma(t_0)}(t_0)\|,t-t_0)=\sqrt{\underline{P}^{-1}}e^{\hat{N}_0\ln\mu+(\frac{\ln\mu}{\tau(t_0,t)}+\tilde{\gamma})(t-t_0)}\sqrt{\overline{P}}\|\tilde{\varepsilon}_{\sigma(t_{0}^+)}(t_{0}^+)\|$ is a class $\mathcal{KL}$ function. This means $\limsup\limits_{t\rightarrow+\infty}\|\tilde{\varepsilon}_{\sigma(t)}(t)\|\leq \epsilon= \sqrt{\underline{P}^{-1}}\iota$. One then concludes via Definition \ref{def_sys} that \eqref{OMAS} achieves ultimately bounded consensus tracking of \eqref{exo_system} under $\widehat{\Phi}^{ind}_k\not\equiv 0$. In particular, if $\overline{h}=0$ and $\widehat{\Phi}_k^{ind}\equiv 0$, then $\overline{\vartheta}=0$ and $\overline{\Theta}=0$, which implies ${\Omega}_k\equiv 0$ and $\tilde{\Omega}_k\equiv 0$, i.e., $\iota=0$. Since $\beta$ is a class $\mathcal{KL}$ function in this case as long as \eqref{thm_ratio} and \eqref{thm_ADT} hold for $j=0$, then one derives that $\lim\limits_{t\rightarrow +\infty}\|\tilde{\varepsilon}_{\sigma(t)}(t)\|=0$,  i.e., \eqref{OMAS} achieves (asymptotic) consensus tracking of \eqref{exo_system}.$\blacksquare$
\end{pf}
\begin{remark}
 Theorem \ref{Thm_1} indicates the consensus tracking problem of the OMAS \eqref{OMAS_compact}-\eqref{Agent_moves} with repelling interactions boils down to the stability problem of the perturbed $M^3D$ system \eqref{switched_sys111} with unstable subsystems. The derived ultimate bound $\epsilon$ comprises two parts respectively corresponding to the perturbation $\tilde{h}_{\sigma(t)}$ and state jump $\tilde{\Phi}_k$.
The condition \eqref{thm_ratio} implies a lower bound $-\frac{{\overline{\gamma}_u-\tilde{\gamma}}}{{\overline{\gamma}_s-\tilde{\gamma}}}$ on the total activation time ratio \cite{878825} between the modes in $\mathcal{P}_s$ and $\mathcal{P}_u$ on each subinterval $[t_k,t_f]$, $k\in\mathbb{N}$. Under non-vanishing perturbations including the state-independent impulses, such a condition along with the piecewise ADT condition \eqref{thm_ADT} ensures that after each agent migration, the stabilizing digraphs $\phi\in\mathcal{P}^+_s$ activate long enough to neutralize the destabilizing digraphs $\phi\in\mathcal{P}^+_u\cup\mathcal{P}^-_{>}$.
\end{remark}

\section{Simulation examples}\label{Sec_4_C}
In this section, a simulation example is presented to verify the obtained results.

\begin{figure}
  \centering
  \includegraphics[width=3in]{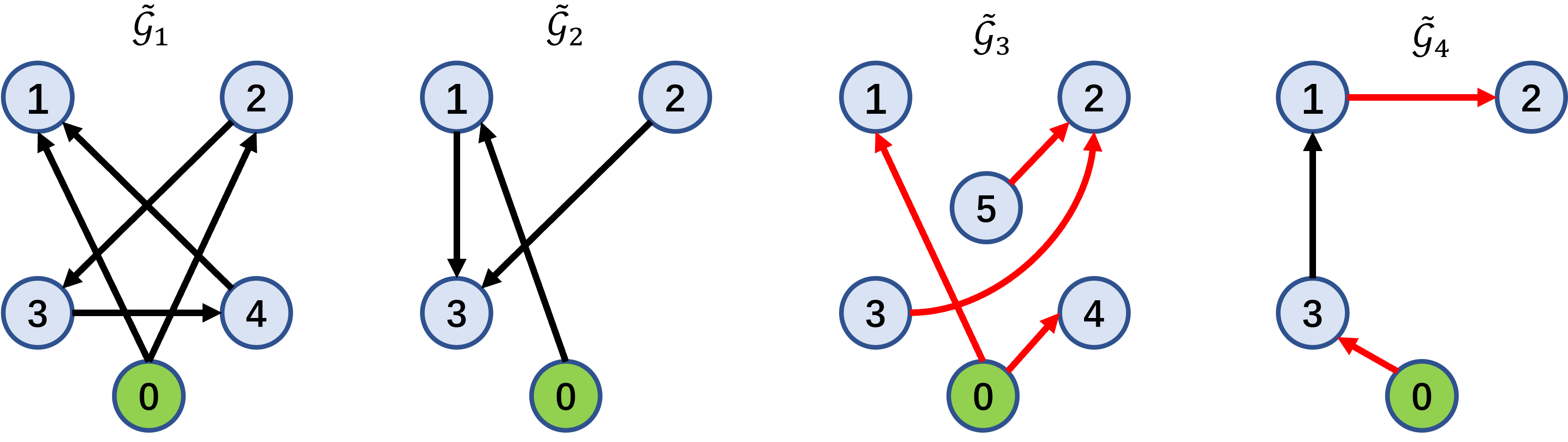}
  \caption{Network topologies $\tilde{\mathcal{G}}_{\phi}$, $\phi=1,2,3,4$. Each blue (green) circle denotes the agent $i\in\mathcal{V}_{\phi}$ (the reference agent 0), each black (red) arrow denotes a positive (negative) edge.
  }\label{fig_top_graph}
\end{figure}

\begin{figure}
  \centering
  \includegraphics[width=3.25in]{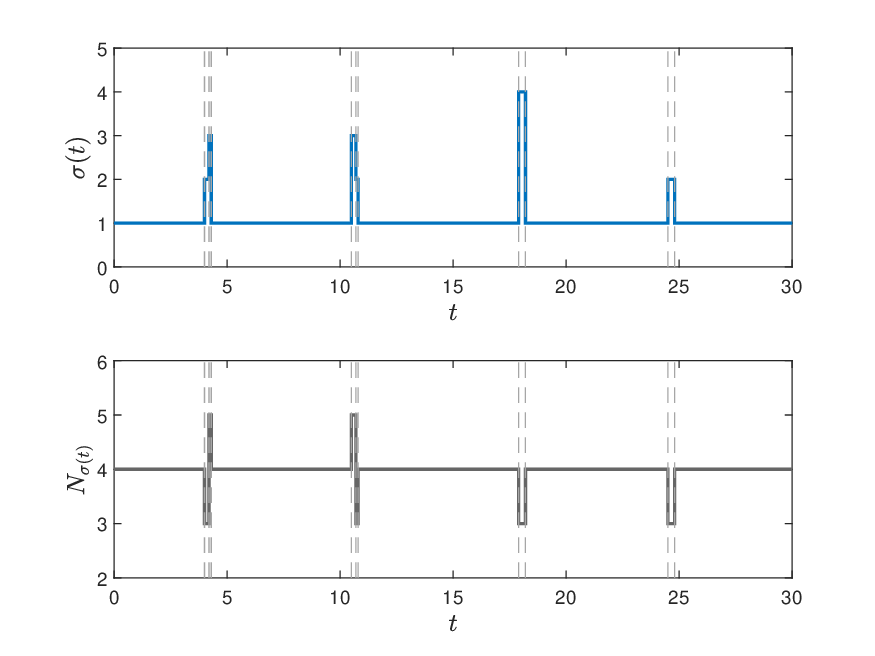}
  \caption{Switching signal $\sigma(t)$ (blue solid line) satisfying \eqref{thm_ratio} and \eqref{thm_ADT}; the agent number $N_{\sigma(t)}$ (gray solid line).}\label{fig_switching}
\end{figure}

Consider the OMAS network topologies $\tilde{\mathcal{G}}_{\phi}$, $\forall \phi\in\mathcal{P}=\{1,2,3,4\}$ depicted in Fig. \ref{fig_top_graph}, from which the corresponding $L_{\phi}$ and $D_{\phi}$ are readily derived for each $\phi\in\mathcal{P}$ as $L_1=[1	~ 0	 ~ 0 ~ -1; 0~  0 ~ 0 ~ 0;0	~ -1 ~	1 ~	0;
0 ~	0 ~	-1 ~ 1]$, $L_2=[0~   0~   0;
    0~   0~   0;
    -1~  -1~   2]$, $L_3=[ 0 ~    0   ~  0 ~    0  ~ 0;
     0 ~    -2 ~   1  ~  0 ~   1;
     0  ~  0  ~ 0 ~    0  ~  0;
          0  ~   0    ~ 0   ~  0   ~  0;
     0 ~   0    ~ 0  ~  0  ~   0]$, $L_4=[1  ~ 0 ~   -1;    1  ~  -1 ~  0;    0  ~0~   0]$,
     $D_1=\mathrm{diag}(1,1,0,0)$, $D_2=\mathrm{diag}(1,0,0)$, $D_3=\mathrm{diag}(-1,0,0,-1,0)$, $D_4=\mathrm{diag}(0,0,-1)$. The OMAS dynamics \eqref{OMAS_compact} is given with
$A=[0~1;-0.2~0.05]$,
 $\varrho=-2.95$. One obtains from the above that $\alpha(\tilde{A}_{1})=-2.925$, $\alpha(\tilde{A}_{2})=0.025$, $\alpha(\tilde{A}_{3})=5.925$, $\alpha(\tilde{A}_{4})=2.975$, which implies $\mathcal{P}^+_s=\mathcal{P}_s=\{1\}$, $\mathcal{P}^+_u=\{2\}$, $\mathcal{P}^+=\mathcal{P}^+_s\cup\mathcal{P}^+_u=\{1,2\}$, $\mathcal{P}^-_{\exists}=\mathcal{P}^-_>=\{3,4\}$, $\mathcal{P}_u=\mathcal{P}^+_u\cup\mathcal{P}^-_>=\{2,3,4\}$. Then, the lower bound for the activation time ratio condition \eqref{thm_ratio} is derived as $13.15$ and the lower bound for the piecewise ADT condition \eqref{thm_ADT} is derived as $2.42$. The switching signal $\sigma(t)$ satisfying \eqref{thm_ratio} and \eqref{thm_ADT} is depicted in the upper subfigure of Fig. \ref{fig_switching} (setting $\hat{N}_0=0$)  while the evolution of the corresponding agent number in $\mathcal{G}_{\sigma(t)}$ is shown in the lower subfigure of Fig. \ref{fig_switching}. The matrices $\Xi_{\sigma(t_k^+),\sigma(t_k^-)}$, $k=1,...,\hat{N}(0,30)$ in \eqref{Agent_moves} can be readily derived from Fig. \ref{fig_top_graph} and Fig. \ref{fig_switching}: $\Xi_{2,1}=[1~     0~     0~     0;
     0~     0~     1~     0;
     0~     0~     0~     1]$, $\Xi_{3,2}=[1~     0~     0;
     0~     0~     0;
     0~     1~     0;
     0~     0~     1;
     0~     0~     0]$, $\Xi_{1,3}=[1~     0~     0~     0~     0;
     0~     1~     0~     0~     0;
     0~     0~     1~     0~     0;
     0~     0~     0~     1~     0]$, $\Xi_{3,1}=[1~     0~     0~     0;
     0~     1~     0~     0;
     0~     0~     1~     0;
     0~     0~     0~     1;
     0~     0~     0~     0]$, $\Xi_{1,2}=[1~     0~     0;
     0~     1~     0;
     0~     0~     0;
     0~     0~     1]$, $\Xi_{2,3}=[1~     0~     0~     0~     0;
     0~     1~     0~     0~     0;
     0~     0~     0~     0~     1]$, $\Xi_{4,1}=[0~     1~     0~     0;
     0~     0~     1~     0;
     0~     0 ~    0 ~    1]$, and $\Xi_{1,4}=[1~         0~     0;
        0~     1~     0; 0~ 0~ 0;
     0~          0~     1]$.
The initial state $\tilde{\xi}_{\sigma(t_0)}(t_0)$, the non-vanishing perturbation $\tilde{h}_{\sigma(t)}(t)$, the state-independent impulse $\tilde{\Phi}_k^{ind}$, and $\widehat{\Xi}_{\sigma(t^+_k),\sigma(t^-_k)}$ for the state-dependent impulse $\tilde{\Phi}_k^{dep}$ are randomly generated (and are thus omitted). In addition, $\overline{\Phi}=0.53$ and $\overline{h}=0.2$.

\begin{figure*}
  \centering
  \subfloat[Trajectories of $\tilde{\xi}_{\sigma(t)}(t)$ and $\tilde{\varepsilon}_{\sigma(t)}(t)$ under non-vanishing $\tilde{h}_{\sigma(t)}(t)$ and state-independent $\tilde{\Phi}_k$.]{\includegraphics[width=3.25in]{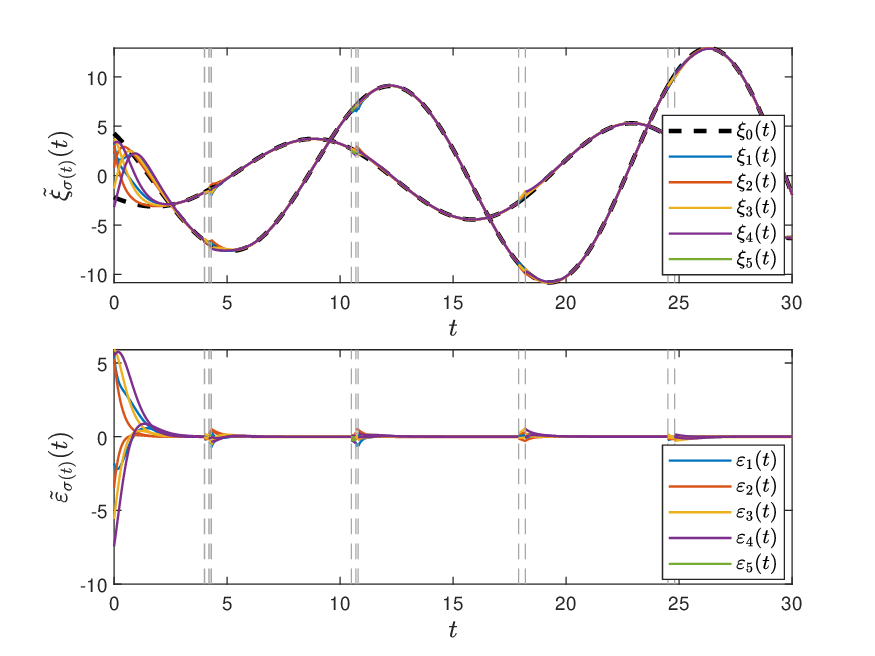}
  \label{fig_state}}
  \hfill
  \centering
  \subfloat[Trajectories of $\tilde{\xi}_{\sigma(t)}(t)$ and $\tilde{\varepsilon}_{\sigma(t)}(t)$ under vanishing $\tilde{h}_{\sigma(t)}(t)$ and state-dependent $\tilde{\Phi}_k$.]{\includegraphics[width=3.25in]{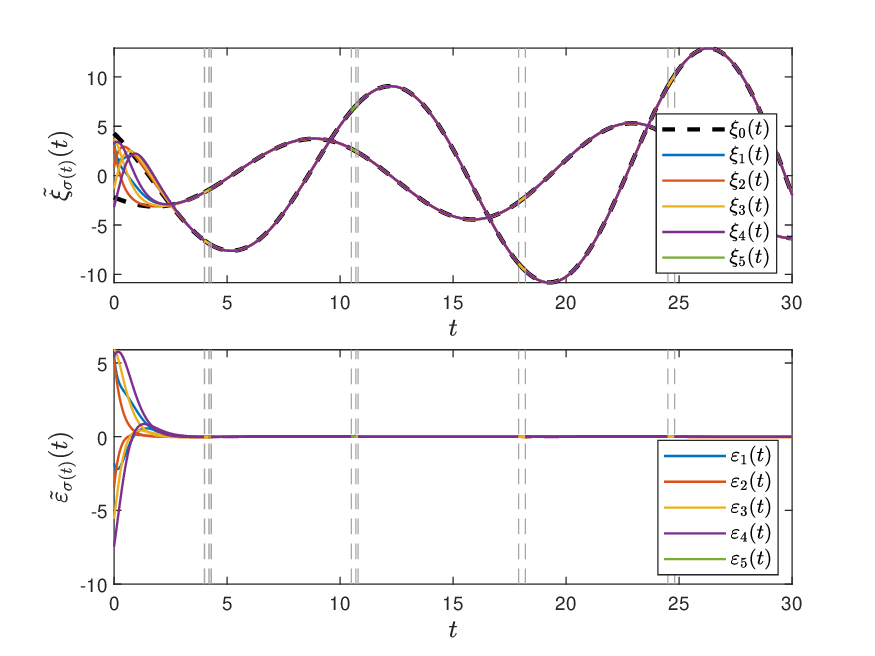}
  \label{fig_state_dep}}
  \caption{Evolutions of the OMAS state $\tilde{\xi}_{\sigma(t)}(t)$ and the tracking error state $\tilde{\varepsilon}_{\sigma(t)}(t)$ under non-vanishing/vanishing perturbations on $[0,30]$. Each legend is for the case of the largest $N_{\sigma(t)}$ over $[0,30]$ ($\max_{\phi}N_{\phi}=5$ in this example).}\label{fig_big}
\end{figure*}

Under the above parameters for dynamics and topologies, we first consider the case of the state-independent impulse $\tilde{\Phi}_k^{ind}\not\equiv 0$ and non-vanishing perturbation $\tilde{h}_{\sigma(t)}(t)$. The resultant trajectories of the OMAS state $\tilde{\xi}_{\sigma(t)}(t)$ and the tracking error $\tilde{\varepsilon}_{\sigma(t)}(t)$ are depicted in Fig. \ref{fig_state}.
  It can be observed that with the non-vanishing property of both the perturbation and state-independent impulse at each migration instant $t_k$, the tracking errors constantly jump but are ultimately bounded while converging to zero asymptotically during each jump-free period, which indicates ultimately bounded consensus tracking is achieved for the OMAS. Then, consider the case of state-dependent impulses and vanishing perturbations, i.e., $\tilde{\Phi}^{ind}_k\equiv 0$,  $\overline{h}=0$. The corresponding trajectories of the OMAS state $\tilde{\xi}_{\sigma(t)}(t)$ and the tracking error $\tilde{\varepsilon}_{\sigma(t)}(t)$ are shown in Fig. \ref{fig_state_dep}. One can observe that owing to the vanishing property of both the perturbation and the state-independent impulse at each $t_k$, the jumps and bumps brought by agent migrations vanish with time goes on as the tracking errors converge to zero, which indicates the OMAS achieves  asymptotic consensus tracking.

\section{Conclusion}\label{Sec_5}
The consensus tracking problem of the OMAS under migration-induced non-vanishing perturbations and repelling antagonistic interactions has been addressed. We show that the OMAS network with repelling Laplacians must be unstable regardless of the network connectivity condition if the repelling interactions outnumber the cooperative ones. In the presence of destabilizing effects brought by both the repelling interactions and the non-vanishing perturbations, we have further extended the stability theory for $M^3D$ systems and applied it to the OMAS to show that ultimately bounded consensus tracking can be reached if the switching conditions of piecewise ADT and activation time ratio are satisfied. In particular, asymptotic consensus tracking can be ensured under vanishing perturbations with weaker switching conditions.

\bibliography{ref1,works_J}

\end{document}